# Tip- and plasmon-enhanced infrared nanoscopy for ultrasensitive molecular characterizations

Y. Luan[1,2], L. McDermott[1], F. Hu[1,2], Z. Fei[1,2]*

[1]Department of Physics and Astronomy, Iowa State University, Ames, Iowa 50011, USA
[2]Ames Laboratory, U.S. Department of Energy, Iowa State University, Ames, Iowa 50011, USA

*Correspondence to: (Z.F.) zfei@iastate.edu.

**ABSTRACT:** We propose a novel method for ultra-sensitive infrared (IR) vibrational spectroscopy of molecules with nanoscale footprints by combining the tip enhancement of the scattering-type scanning near-field optical microscope (s-SNOM) and the plasmon enhancement of the breathing-mode (BM) plasmon resonances of graphene nanodisks (GNDs). To demonstrate that, we developed a quantitative model that is capable of computing accurately the s-SNOM signals of nanoscale samples. With our modeling, we show that the s-SNOM tip can effectively excite gate-tunable BM plasmonic resonances in GNDs with strong field enhancement and sensitive dependence on the size of GND. Moreover, we demonstrate that the intense electric field of tip-excited plasmonic BMs can strongly enhance the IR vibrational modes of molecules. As a result, IR vibrational signatures of individual molecular particles with sizes down to 1-2 nm can be readily observable by s-SNOM. Our study sheds light on future ultra-sensitive IR biosensing that takes advantage of both the tip and plasmon enhancement.

## I. INTRODUCTION

Graphene plasmon polaritons are collective oscillations of Dirac fermions in graphene covering a broad spectral range from terahertz to infrared [1-7]. With both imaging and spectroscopy techniques [8-11], graphene plasmons have been extensively studied in recent years. Many superior characteristics have been discovered, such as electrical tunability [8-14], strong confinement [8-17], long lifetime [17,18], and high environmental sensitivity [17-19]. These unique properties lead to many technological innovations, among which plasmon-enhanced infrared spectroscopy (PEIS) is possibly the most promising one. Indeed, recent experimental studies demonstrated that graphene plasmon resonances can enhance the infrared (IR) vibrational modes of molecules or polymers [20-23]. In nearly all these studies, dipole-mode (DM) plasmon resonances of graphene nanostructures (GNS) were excited directly by far-field IR beams. Despite the success of PEIS demonstrated with GNS, the sensitivity is still too low to probe individual molecules or bio-particles, which is partly due to the diffraction-limited IR beam with a large beam size, and partly due to the relatively weak field enhancement of the DM plasmon resonances.

In order to increase the sensitivity for nanoscale IR characterizations of small molecules, stronger field enhancement and higher spatial resolution are necessary. For that purpose, we propose to take advantage of the breathing-mode (BM) plasmonic resonances [24-26] in graphene nanodisks (GNDs) that could induce even stronger field enhancement due to the circular symmetry. Unlike DM plasmon resonances, the plasmonic BMs have zero net polarization, so they are considered dark modes that cannot be excited directly by far-field optical beams [24]. To excite and probe these BMs, the scattering-type scanning near-field optical microscope (s-SNOM) [Fig. 1(a)] has been proven to be an effective tool [25]. When illuminated by a *p*-polarized laser, the sharp metallic s-SNOM tip acts as a nano-antenna that is perfect for exciting the circular-symmetric BMs in GNDs [Fig. 1(b)]. Moreover, the tip can further enhance the plasmon field due to the so-called "lightning-rod effect". As discussed below, the two-fold enhancement by both the tip and BM plasmons is the key to the ultra-high sensitivity of the proposed method here. Beside mode excitation and field enhancement, the s-SNOM tip also plays the role of plasmon field detection. Indeed, the radiation signal from the polarized s-SNOM tip collected by the detector offers an accurate measurement of the plasmon field right underneath the tip. Such a capability of tip excitation & detection of s-SNOM has been demonstrated in the studies of graphene plasmons [9-11] and other polaritonic modes in two-dimensional (2D) materials [27-36]. Furthermore, s-SNOM enables nanoscale Fourier-transform IR spectroscopy (nano-FTIR) when coupled with a broadband IR laser, which is suitable for vibrational mode fingerprinting of nanoscale materials and biomolecules [37-40].

In this paper, we demonstrate with quantitative modeling that s-SNOM, when coupled with BM plasmon resonances of GNDs, is perfect for applications in ultra-sensitive PEIS. In the following sections, we will first introduce our quantitative s-SNOM model with an elongated tip shape, which is essential for accurately modeling the s-SNOM signals. We will then discuss the general characteristics of the tip-excited BM plasmons of GNDs with modeled nano-IR spectra and plasmon field patterns. Finally, we will demonstrate the ultra-high sensitivity of s-SNOM to vibrational modes of molecules due to the field enhancement by both the s-SNOM tip and the BM plasmon resonances.

## II. RESULTS AND DISCUSSION
### A. Quantitative s-SNOM model for nanoscale samples.

To model quantitatively the s-SNOM signals, we construct a numerical model based on the finite-element commercial solver Comsol Multiphysics. Note that there were several earlier works about quantitative models of s-SNOM [41-44], which typically assume infinite lateral sizes of samples for simplicity in computation. As discussed below, the Comsol-based s-SNOM model introduced here can be conveniently tailored for nanoscale samples with small dimensions. Depending on the symmetry of the samples in consideration, we choose either 2D axisymmetric or 3D Comsol models. The former is more efficient in computation and is more suitable for the modeling of BM plasmons of GNDs, and the latter can in principle be used to model sample structures with arbitrary geometries. The results with both models are consistent with each other. The results shown in this work are mainly obtained with the 2D axisymmetric model. Typical computation time for one complete spectrum with this model is only about 1 hour or less when using a common desktop workstation.

As illustrated in Fig. 1(a), the s-SNOM tip is modeled as the conducting spheroid with a length of $L$ and a radius of curvature at the tip apex of $a$, where $a$ is set to be about 25 nm according to the manufacturer and $L$ is set to be 1200 nm to best fit the s-SNOM data of standard reference materials ($SiO_2$ and graphene, see Supplemental Material [45]). With the 2D axisymmetric model, we evaluate numerically the total radiating dipole ($p_z$) of the tip, which is roughly proportional to the scattering field $E_s$ measured by the s-SNOM [9,44]. Due to the elongated shape of the tip, we neglect horizontal dipole moments of the tip ($p_x$ and $p_y$) that are orders of magnitude weaker than $p_z$. Note that the s-SNOM is commonly built based on a tapping-mode atomic force microscope (AFM). The tapping amplitude ($A$) is typically in the order of tens of nanometers ($A$ set to be 40 nm in the current work). As a result, the scattering signal is naturally modulated due to the tapping of the tip. Demodulation of the scattering signal at $n^{th}$ harmonics ($n \geq 2$) of the tapping frequency can significantly suppress the background signals and thus capturing genuine near-field responses [43,44,46]. In our modeling, we mimic the modulation & demodulation data acquisition process of s-SNOM, and obtain the $n^{th}$ harmonics of the near-field amplitude signal ($n = 3$ in the current work) by calculating $p_z$ of the tip at multiple tip-sample separations [45]. For the purpose of quantitative comparison, we normalized the s-SNOM amplitude signal to that of silicon due to the flat spectral response of silicon in the mid-IR region. In all the calculated nano-IR spectra shown below, we plot the normalized near-field amplitude signal ($s$) of s-SNOM. The signal to noise ratio of an optimally-aligned s-SNOM is up to ~100 when measuring silicon, so any spectroscopic features with sizes larger than ~1% of the signal of silicon could be observable in practical s-SNOM experiments with optimized alignment.

Graphene is modeled as a 3D metal with a thickness of $t_g = 0.34$ nm, and the corresponding 3D optical conductivity is set to be $\sigma_{3D} = \sigma_{2D}/t_g$, where the 2D conductivity of graphene ($\sigma_{2D}$) is obtained with the random-phase approximation method as detailed in Ref. [9]. The two parameters of graphene are the Fermi energy ($E_F$) and the phenomenological scattering rate $\gamma$. The latter is defined as the ratio between the charge scattering energy of graphene and the IR excitation energy. Throughout the main text, we set $\gamma$ to be 0.1 that is consistent with previous experimental studies [9,10]. As a test, we show in Fig. S1 the near-field spectra of $SiO_2$ and graphene on $SiO_2$. The quantitative consistency between the modeling results with previous experimental data [9,43] justifies the validity of our model. Detailed discussions about the test modeling results are given in the Supplemental Material [45].

**B. Breathing-mode plasmon resonances.**

With the quantitative model, we first study the spectroscopic responses of GNDs on a $CaF_2$ substrate. The optical constants of $CaF_2$ that we used in the modeling are adoped from previous literature [47]. As shown in Fig. S2, $CaF_2$ has a relatively flat mid-IR response in a wide spectral range above 60 meV. The spectral response below 60 meV is dominated by optical phonons of $CaF_2$. As a comparison, the commonly-used $SiO_2$ substrate has three optical resonances centered at around 50 meV, 100 meV, and 140 meV, respectively (Fig. S2). Therefore, $CaF_2$ serves as a better substrate compared to $SiO_2$ for revealing the pure plasmonic responses of graphene in the wide mid-IR region, as confirmed by previous experimental studies [21,25].

Figure 2(a) presents the calculated nano-IR spectra of GNDs with various radius ($r$) using our s-SNOM model. For all the calculations, the tip is located at the center of the GNDs. The $E_F$ of graphene is set to be 0.2 eV. The key features in the calculated s-SNOM spectra are the resonance peaks (marked with arrows). The number of resonance peaks within the considered spectral region (60 – 160 meV) increases with the GND radius. As shown in the top panel of Fig. 2(a), there is clearly one dominant peak (marked with the black arrow) in the near-field spectrum of the GND with $r$ = 50 nm. As $r$ increases, the dominant peak (black arrow) shifts to lower energies and the 2nd (blue arrow) and 3rd (red arrow) peaks emerge at higher energies. More peaks emerge in even larger GNDs (e.g. $r$ = 300 - 500 nm). Nevertheless, the higher-order peaks are much weaker and closer to each other, so they become less distinguishable. They can be seen more clearly in cleaner samples with a lower scattering rate (see Fig. S5 in the Supplemental Material [45]). Note that we used a logarithmic scale for the normalized IR amplitude in Fig. 2.

In addition to the size dependence, we also explored the doping dependence of these resonance modes. In Fig. 2(b), we plot the calculated nano-IR spectra of a 300-nm-radius GND with various $E_F$. Multiple resonance peaks (marked with arrows) are clearly seen in the spectra with $E_F \geq 0.1$ eV and the peak locations shift to higher energies as $E_F$ increases. In the case of $E_F = 0.05$ eV [top panel of Fig. 2(b)], all the resonance peaks are close to or below 60 meV, so they are not fully shown in the field of view. Instead, the dominant feature here is originated from the interband transitions with onset energy at $2E_F$ (marked with vertical dashed line) [48-50].

The resonance modes of GNDs shown in Fig. 2 are originated from the BM plasmonic resonances. To demonstrate that, we computed the $z$-component electrical field ($E_z$) maps of GNDs at the first (black arrow), second (blue arrow) and third (red arrow) resonance locations of the GND with $r$ = 300 nm and $E_F$ = 0.3 eV [see Fig. 2(b)]. The corresponding out-of-plane ($x$-$z$ plane, left panels) and in-plane ($x$-$y$ plane, right panels) $E_z$ maps are shown in Fig. 3, where one can see that the spatial field patterns correspond to BMs with different orders ($n$ = 0, 1, 2 …). The order index assignment will be discussed in detail below. To verify the plasmonic origin of these BMs, we plot in Fig. 4(a) and Fig. 4(b) the resonant peak energies ($E_p$, data points) versus $1/r$ and $E_F$ respectively based on the nano-IR spectra in Fig. 2. In both cases, $E_p$ increase with increasing $1/r$ or $E_F$ for all three resonance modes marked with arrows. Such dependence behaviors are fully consistent with the dispersion properties of graphene plasmons.

Under long-wavelength approximation, the dispersion relation of graphene plasmons is described by

$$q_p \approx \frac{i\varepsilon_0(1+\varepsilon_s)e}{\hbar} \frac{E}{\sigma_{2D}}, \quad [1]$$

where $q_p$ is the plasmon wavevector, $\varepsilon_s$ is the dielectric function of the $CaF_2$ substrate [45]. Under Drude approximation, the 2D conductivity of graphene can be written as

$$\sigma_{2D} \approx \frac{ie^2}{\pi\hbar} \frac{E_F}{E(1+i\gamma)}. \quad [2]$$

Based on Eq. (1) and Eq. (2), we have the plasmon energy

$$E_p^2 \approx \frac{e}{\pi} \frac{q_p E_F}{\varepsilon_0(1+\varepsilon_s)}. \quad [3]$$

Therefore, in a relatively flat dielectric environment, $E_p$ scales with the square root of $q_p$ and $E_F$. Similar to the DM plasmon resonance modes of GNRs, the characteristic mode wavevector $q_p$ of a breathing mode in GNDs is also determined by the size of the disk. The mode equation can be written as:

$$q_p r + \Phi_R = n\pi, \qquad [4]$$

where $\Phi_R \approx -0.75\pi$ is the anomalous phase shift upon reflection off the edge of graphene [12,51,52]. Equation 4 indicates that $q_p$ is proportional to $1/r$ for all the resonance modes. Therefore, the $1/r$ and $E_F$ dependence relations of plasmon resonance energy shown in Fig. 4 can be fully understood by Eq. (3) and Eq. (4). For the purpose of quantitative comparison, we plot in Fig. 4 calculated $1/r$ and $E_F$ dependence relations (curves) of graphene plasmons based on Eq. (3) and Eq. (4). From Fig. 4, one can see that the resonance energies (data points) extracted from nano-IR spectra (Fig. 2) are generally consistent with the analytical curves, indicating that the observed resonance modes are indeed due to graphene plasmons. There are slight deviations (5-10%) in the $n = 0$ mode in Fig. 4(b) in small GNDs ($r = 50$ and $100$ nm). This is mainly due to the impact of the metallic tip on the plasmon resonance energy of GND, which will have notable effects when the size of GND is very small. Similar phenomena have been observed experimentally in previous studies [53,54], where the strong coupling between graphene and adjacent metal layer leads to the formation of acoustic plasmons. More discussions about the tip-induced modification of plasmon resonance energy are given in the Supplemental Material [45].

### C. Tip- and plasmon-enhanced vibrational modes of molecules.

Now we wish to study the enhancement effects of molecular vibrational modes by the tip-excited BM plasmonic resonances. We consider a disk-shaped particle consisting of pentacene molecules sitting at the center of the GND [see Fig. 1(a)]. Here the pentacene molecules are standing upright, which is energetically favored according to previous literature [55]. The thickness of the molecular disk is set to be ~1.6 nm which is roughly the length of the pentacene molecule. Other particle shapes are discussed in the Supplemental Material [45]. In our model, we use the vibrational mode of pentacene at 112 meV as an example. The enhancement effects on other vibrational modes of pentacene or other types of molecules are qualitatively similar. The vibrational mode is modeled by the following LT oscillator:

$$\varepsilon(E) = \varepsilon_\infty + \frac{S E_{TO}^2}{E_{TO}^2 - E^2 - iEE_\Gamma}, \qquad [7]$$

where $\varepsilon_\infty = 2.981$, $E_{TO} \approx 112$ meV, $S = 0.0102$, and $E_\Gamma = 0.43$ meV according to Ref. [56].

The modeling results are shown in Fig. 5(a), where we plot the calculated nano-IR spectra of a molecular particle on GND (red curve), bare GND (black curve), and the bare molecular particle (blue curve). The radius of the molecular disk ($\rho$) is set to be 5 nm. The radius and Fermi energy of GND are set to be 100 nm and 0.26 eV, respectively. Under these settings, one can see a strong and broad resonance peak centered at 107 meV in the spectrum of bare GND [black curve in Fig. 5(a)], which is attributed to the $n = 0$ BM plasmon resonance. The bare molecule spectrum [blue curve in Fig. 5(a)], on the other hand, is extremely weak and is not seen on the same scale. A zoom-in view [inset of Fig. 5(a)] of the bare molecule spectrum shows the peak feature due to the molecule vibrational mode at about 112 meV. Nevertheless, the peak feature is only less than 0.001 when normalized to silicon [we multiply the amplitude signal by 150 times in the inset of Fig. 5(a)], so it is impossible for observation in practical measurements. Adding the GND underneath molecules significantly enhances the vibrational mode feature, which appears to be a sharp dip on top of the broad plasmon resonance. The dip feature has a size of 0.3 when normalized to silicon, so it can be easily detectable by s-SNOM. The formation of the dip feature instead of the peak feature is originated from the interference between the molecular vibrational mode with the plasmon resonance mode, which was described as the so-called "plasmon-induced transparency" in previous far-field studies [57,58]. From Fig. 5(a), we also notice variations of the plasmon resonance energy when comparing the spectrum of molecules/GND (red curve) to that of bare GND (black curve). This is mainly due to the enlarged tip-GND separation after adding the molecular particle (see Supplemental Material [45]). Indeed, if intentionally increasing the tip-GND separation by 1.6 nm (the height of molecules), the plasmon resonance energy of bare GND [black dashed curve in Fig. 5(a)] matches better that of molecules on GND [red curve in Fig. 5(a)].

Now we wish to explore the sensitivity limit of the proposed method for probing extremely small molecular particles. As shown in Fig. 5(b), we plot the calculated s-SNOM spectra of molecular disks with various radii ($\rho$) on GNDs. Clearly, the dip feature due to the vibrational mode can be visualized in molecules with a disk radius down to 1 nm, corresponding to a volume of about 5 nm$^3$. The relative size of the dip feature is close to 0.05 when normalized to silicon, so it is clearly detectable by s-SNOM. Such a high sensitivity is impossible with traditional diffraction-limited IR spectroscopy or nano-FTIR without GND. The highest sensitivity demonstrated so far with s-SNOM alone is reported in Ref. [59], where the vibrational modes of biomolecules with a volume ~ 900 nm$^3$ (sphere with a diameter of 12 nm) were observed. Therefore, with additional enhancement by the BM plasmon resonance of GND, the probing sensitivity increases dramatically.

The capability of ultra-sensitive detection is originated from the strong field enhancement from both the metal tip and the BM plasmon resonance of GND. To demonstrate that, in Fig. 6, we compare the simulated $E_z$ field maps with four different settings: (1) metal tip with GND [Fig. 6(a)], (2) silicon tip with GND [Fig. 6(b)], (3) metal tip without GND [Fig. 6(c)], and (4) silicon tip without GND [Fig. 6(d)]. As a general example, we set the tip-sample distance to be 5 nm, the excitation energy to be 112 meV that matches the resonance energy of GND with $E_F$ = 0.26 eV. The metal here is set to be platinum that is commonly used for the coating material of s-SNOM tips. The purpose of comparing the metal tip with a silicon tip is to evaluate the field enhancement effects due to the metallic tip of s-SNOM (the field enhancement effects due to a silicon tip is limited). From the simulations (Fig. 6), we find that the metal tip with GND [Fig. 6(a)] produces the highest field amplitude ($|E_z|$) underneath the tip, which is ~5 times higher than that with a silicon tip and GND [Fig. 6(b)], ~10 times higher than that with metal tip but no GND [Fig. 6(c)], ~24 times higher than that with bare silicon tip [Fig. 6(d)], and ~225 times compared to that of the far-field excitation field. Therefore, the strong field enhancement is originated from both the metallic tip and the BM plasmon resonance of GND. Note that the $E_z$ field values here are mainly for a qualitative understanding of the enhancement effects. To understand the exact size or shape of the spectroscopic feature of the vibrational mode in Fig. 5, one needs to consider the interference between the molecule vibrational mode and plasmon resonance mode of GND as well as the complicated tip modulation & demodulation signal acquisition processes.

Finally, we wish to discuss several considerations related to practical applications of the proposed method. First, the plasmon enhancement effect described above requires matching between the plasmon resonance energy with the vibrational mode of molecules. Nevertheless, the match is not stringent due to the broad resonance of the $n = 0$ BM plasmon resonance of GND. As shown in Fig. 5(c,d), we plot the calculated nano-IR spectra with variations of $E_F$ [0.22 – 0.30 eV, see Fig. 5(c)] and the GND radius [80-110 nm, see Fig. 5(d)]. The spectroscopic feature due to the molecule vibrational mode can be seen clearly in all the spectra indicating that the requirements of selecting $E_F$ or GND radius for experimental measurements are not stringent. Second, all the modeling results discussed above assume that the tip and molecule are right at the center of GND. In fact, even the tip and molecules are slightly off the center, field enhancement effects due to both the tip and GND are still strong. For example, when the tip is 20 nm and 40 nm away from the center of GND with $r$ = 100 nm, the $E_z$ field amplitude only drops by 12% and 35%, respectively (Fig. S6). Third, for practical experiments, we propose to fabricate electrically-connected GND arrays with chemical-vapor-deposited graphene (Fig. S7, similar to the device shown in Ref. [60]). Such devices are economically feasible for fabrications based on common lithography methods. These densely distributed GND arrays act as a testbed, where molecules or other nanoparticles of interest can be dispersed on top. After locating the molecular particle on GNDs with AFM topography scanning, one can perform nano-FTIR spectroscopy after optimizing the signals of the molecular vibrational modes with electrical gating.

## III. CONCLUSION

In conclusion, we demonstrate through vigorous modeling a novel method for ultra-sensitive IR vibrational spectroscopy by taking advantage of the field enhancement of both a sharp metallic tip and the BM plasmon resonances of GNDs. Due to the two-fold enhancement, the electric field between the tip apex and the GND can be orders of magnitude higher than that of the free-space excitation field. As a result, an ultra-high sensitivity to molecular vibrational modes could be achieved, which enables the detection of molecular particles with a size down to 1-2 nm. Note that the field enhancement method proposed here is not only suitable for probing IR vibrational modes of molecules or phonon modes of crystalline nanoparticles, it can also be used to probe high-field nonlinear effects [61] of nanoparticles in the mid-IR region. Our work paves the way for practical applications of tip- and plasmon-enhanced IR spectroscopy for ultra-sensitive detections and characterizations of small molecules and nanoparticles.

**ACKNOWLEDGMENTS**
Work done at Ames Lab was supported by the U.S. Department of Energy, Office of Basic Energy Science, Division of Materials Sciences and Engineering. Ames Laboratory is operated for the U.S. Department of Energy by Iowa State University under Contract No. DE-AC02-07CH11358.

**Figures**

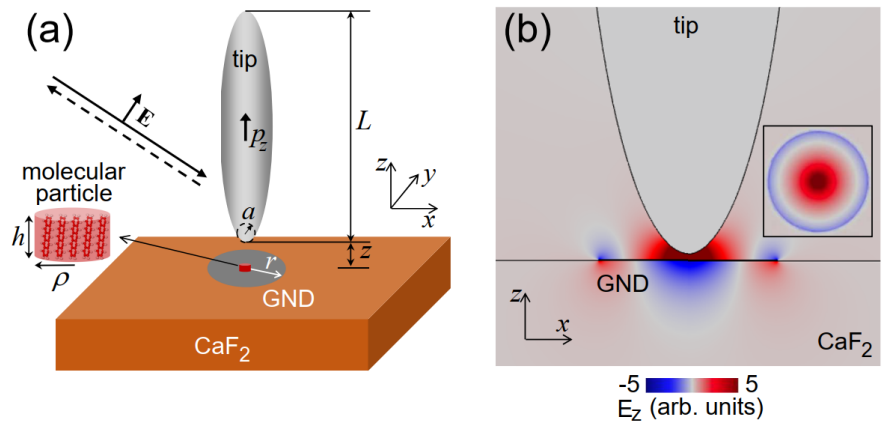

**Fig. 1.** (a), Illustration of the quantitative model of s-SNOM on a molecule/GND sample. The GND has a radius of $r$. The molecular particle has a disk shape with a radius of $\rho$ and thickness of $h$. The s-SNOM tip is approximated as a spheroid with a length of $L$ and a tip-apex radius of $a$. (b) Modeled $E_z$ field map at the $x$-$z$ plane revealing tip-excited plasmonic breathing mode (BM) of a GND. Inset plots the in-plane ($x$-$y$ plane) $E_z$ field map revealing the plasmonic BM with a circular symmetry.

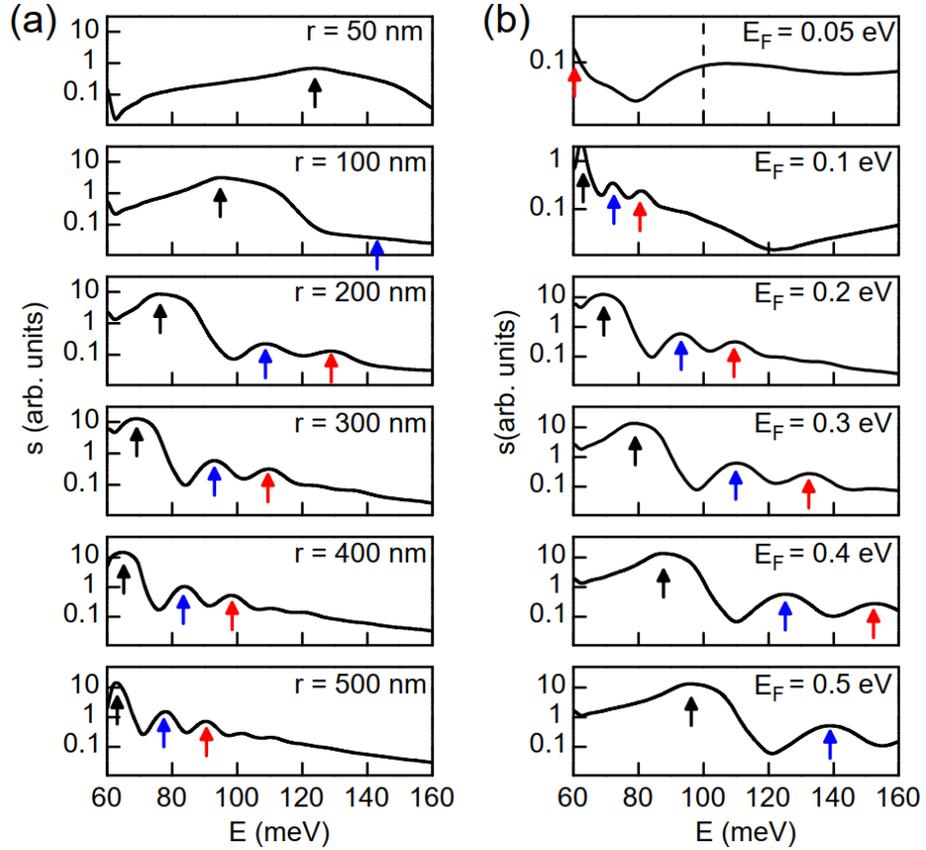

**Fig. 2.** (a) Modeled nano-IR spectra of GNDs with various disk radii on a CaF$_2$ substrate. The Fermi energy ($E_F$) of graphene is set to be 0.2 eV. (b) Calculated nano-IR spectra of GNDs on a CaF$_2$ substrate with varying $E_F$. The GND radius ($r$) is set to be 300 nm. The vertical dashed line in the top panel of (b) marks the onset of interband transitions at $E = 2E_F$. In all panels, a logarithmic scale is used in the y-axis to better visualize all resonance peaks, and the IR amplitude is normalized to that of a bare silicon substrate.

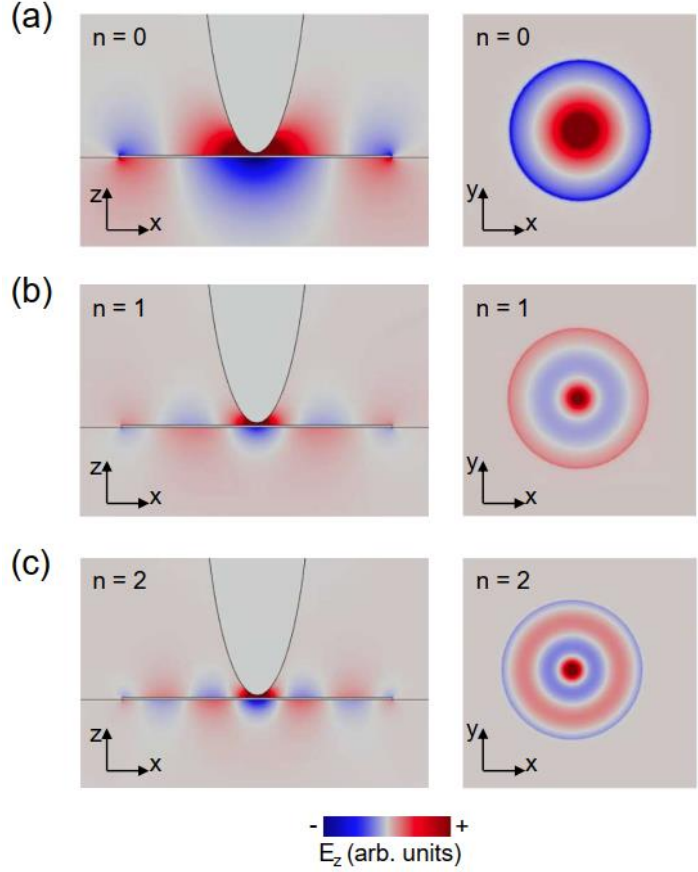

**Fig. 3.** Modeled $E_z$ field maps of the $n = 0, 1, 2$ [from (a) to (c)] BM plasmons of the GND with $r = 300$ nm and $E_F = 0.3$ eV. The resonance energies ($E_p$) of the $n = 0, 1, 2$ modes are about 77 meV, 109 meV, and 133 meV, respectively [see Fig. 2(b)]. The left and right panels are for x-z and x-y planes, respectively. The field patterns of BM plasmons in GNDs with other radius and Fermi energies share similar characteristics.

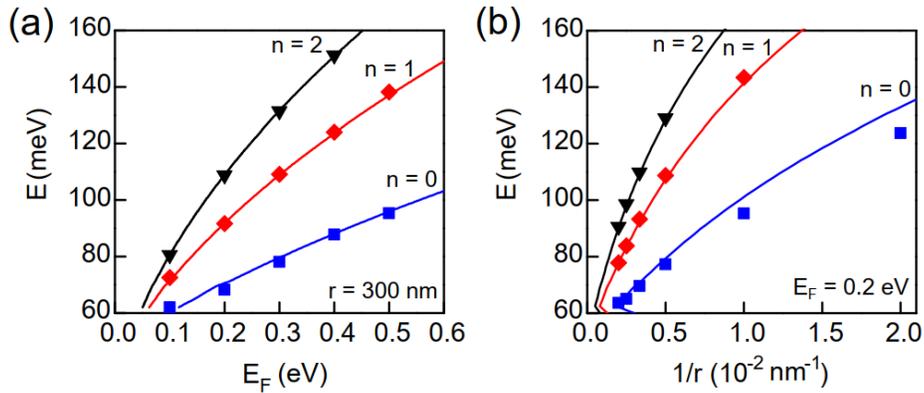

**Fig. 4.** (a) Dependence relationship of the resonance energy $E_p$ on $E_F$ for the $n = 0, 1, 2$ resonance modes. (b) Dependence relationship of $E_p$ on $1/r$ for the $n = 0, 1, 2$ resonance modes. Data points are extracted from the modeled nano-IR spectra in Fig. 2. The color curves were calculated with analytical equations [Eq. (3) and Eq. (4)].

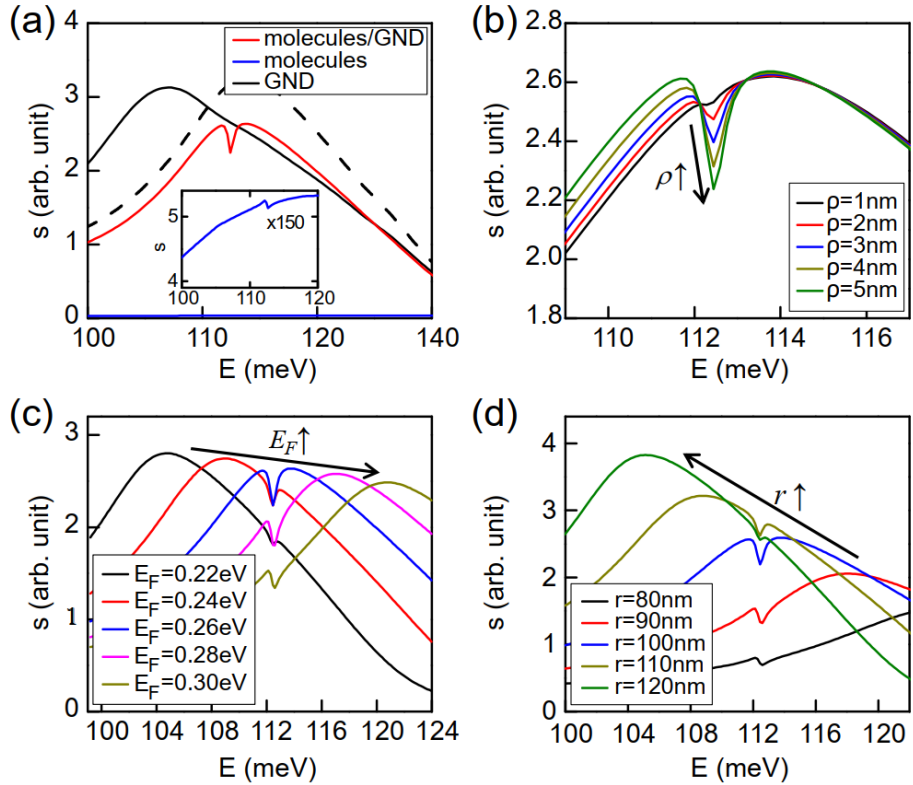

**Fig. 5.** (a) Modeled nano-IR spectra of bare GND (black curve), bare molecules (blue curve), and molecules on GND (red curve). The black dashed curve is for the spectrum of bare GND when the tip-sample distance increases by 1.6 nm (thickness of molecular disk). Inset plots a zoom-in view (amplitude × 150) of the spectrum of bare molecules. Here, Fermi energy is $E_F = 0.26$ eV, GND radius is $r = 100$ nm, and molecule disk radius is $\rho = 5$ nm. (b) Modeled nano-IR spectra of a molecular disk on GND with various molecule disk radii. Here, $E_F = 0.26$ eV and $r = 100$ nm. (c) Modeled nano-IR spectra of molecules with various $E_F$ of GND. Here, $r = 100$ nm and $\rho = 5$ nm. (d) Modeled nano-IR spectra of molecules on GND with various radii of GND. Here, $E_F = 0.26$ and $\rho = 5$ nm. In all panels, the IR amplitude is normalized to that of a bare silicon substrate.

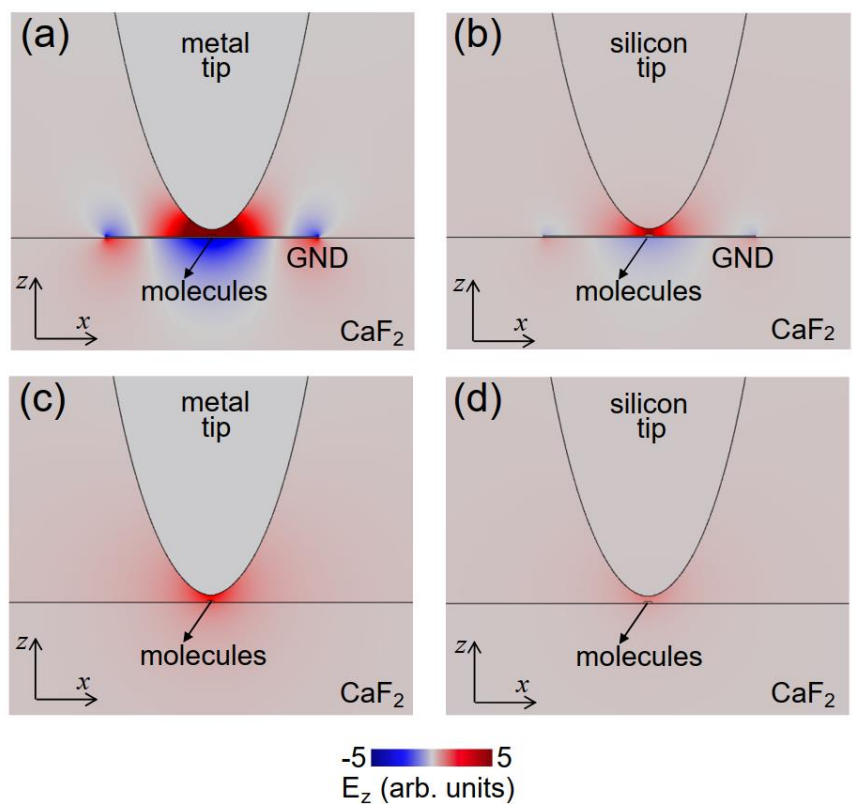

**Fig. 6.** Modeled $E_z$ field maps of the tip-sample system with 4 different settings: (a) metal tip with GND, (b) silicon tip with GND, (c) metal tip without GND, and (d) silicon tip without GND.

**Supplemental Material**
**Tip- and plasmon-enhanced infrared nanoscopy for ultrasensitive molecular characterizations**

Y. Luan[1,2], L. McDermott[1], F. Hu[1,2], Z. Fei[1,2]*

[1]Department of Physics and Astronomy, Iowa State University, Ames, Iowa 50011, USA
[2]Ames Laboratory, U.S. Department of Energy, Iowa State University, Ames, Iowa 50011, USA

*Correspondence to: (Z.F.) zfei@iastate.edu.

**List of contents**

1. Additional information about the s-SNOM model

2. Testing the s-SNOM model

3. The $CaF_2$ substrate

4. Effects of the metallic tip on plasmon resonance energy

5. Comparison of different molecular-particle shapes

    Figures S1 – S7

## 1. Additional information about the s-SNOM model

As discussed in the main text, we used mainly the axisymmetric 2D model to calculate the near-field signals of scattering-type scanning near-field optical microscope (s-SNOM) due to its higher efficiency than the 3D model. In the model, only the $z$-axis polarization of the tip is accounted for due to the axisymmetric. Considering the elongated shape of the s-SNOM tip, in-plane polarization is orders of magnitude smaller than the z-axis polarization, so it won't affect significantly the computation results. Besides, we considered a uniform far-field excitation field along the $z$ direction for the tip. In practical s-SNOM experiments with a side illumination configuration [see Fig. 1(a) in the main text], the background excitation field of the tip is formed by both the incident beam and the reflected beam off the sample surface. The interference between the incident and reflected beams will in principle causes spatial variations of the overall excitation field background, but the field variations are relatively small due to the weak reflection off the CaF$_2$ substrate (reflection coefficient ~ 20%). Moreover, the spatial variation due to interference is in the scale of the IR wavelength (~ 10 µm). Considering the sample (graphene nanodisk and molecule) and the tip apex (radius ~ 20 nm and length << 1 µm) is orders of magnitude smaller than the IR wavelength, the excitation field close to the tip apex and sample is approximately uniform.

As discussed in the main text, the scattering-type scanning near-field optical microscope (s-SNOM) is typically built based on a tapping-mode atomic force microscope (AFM). Due to the tip-tapping, the s-SNOM signal is modulated. Demodulating the s-SNOM signal at higher harmonics of the tapping frequency could generate genuine near-field signals. Our quantitative model computes accurately the s-SNOM signals by considering the tip modulation & demodulation acquisition process, which can be summarized mathematically by following two equations:

$$s_n(\omega) = \int_0^T e^{in\Omega T} E_{rad}(\omega, \Omega, t) dt, \quad [S1]$$

$$d = d_0 + A[1 + \sin(\Omega t)]. \quad [S2]$$

In the above two equations, $S_n$ is the $n^{th}$ harmonics of the near-field amplitude signal ($n = 1, 2, 3,…$) obtained by demodulating the near-field signal at $n$ times of the tapping frequency ($\Omega$), $E_{rad}$ is the radiation signal of the polarized tip that is roughly proportional to the total $z$-component dipole moment ($p_z$) of the tip: $E_{rad} \propto p_z$, $d$ is the tip-sample distance with a minimal value of $d_0$, and $A$ is the tapping amplitude of the tip. In our calculations shown in the main text, we consider $d_0 = 1$ nm and $A = 40$ nm. In practical s-SNOM experiments, $d_0$ is about 1 nm or below depending on sample surface conditions, and $A$ could vary dramatically in different experiments but typically in the order of tens of nanometers.

## 2. Testing the s-SNOM model

As introduced in detail in the main text, we developed the quantitative model based on Comsol Multiphysics. Before applying the model in calculations of the breathing-mode (BM) plasmon resonances of graphene nanodisks (GNDs) and vibrational modes of molecules, we first tested the model in a variety of well-studied materials, such as SiO$_2$ and graphene on SiO$_2$. In Fig. S1, we plot the calculated nano-IR spectra of SiO$_2$ with different thicknesses and graphene on SiO$_2$. The main feature in all these spectra is the surface phonon resonance of SiO$_2$ peaked at about 140 meV. As shown in these spectra, the phonon intensity shows systematic evolution with the SiO$_2$ thickness [Fig. S1(a)]. When adding graphene on top of SiO$_2$, the phonon resonance can be significantly enhanced when graphene is significantly doped, and the enhancement is tunable by adjusting the Fermi energy ($E_F$) of graphene [Fig. S1(b)]. This is due to the coupling between SiO$_2$ phonons with graphene plasmons (Ref. 9 in the main text). The modeling results in Fig. S1 are consistent with experimental data in previous works (Refs. 9 and 43 in the main text), which verifies the validity of our s-SNOM model in computing both the bulk and two-dimensional (2D) materials.

### 3. The CaF$_2$ substrate

The substrate that we chose for the studies of GNDs and molecular vibrational modes is Calcium Fluoride (CaF$_2$). The optical constants that we used for modeling CaF$_2$ are plotted in Fig. S2(a), which were adopted from previous literature (Ref. 47 in the main text). In Fig. S2(b), we plot the calculated nano-IR spectra of both the CaF$_2$ and SiO$_2$ substrates. From both Fig. S2(a,b), one can clearly see that there are strong phonon resonances below 60 meV in CaF$_2$. Nevertheless, the spectral response is quite flat due to the absence of phonons above 60 meV. As a comparison, the commonly used SiO$_2$ substrate has more widely distributed phonons centered at 50 meV, 100 meV, and 140 meV, respectively [Fig. S2(b)]. Therefore, CaF$_2$ is a better substrate to use for the studies of pure plasmon resonance modes of graphene and graphene nanostructures in the wide mid-IR spectral range.

### 4. Effects of the metallic tip on plasmon resonance energy

In addition to the strong enhancement of the plasmon resonance of GND, we found that the metallic tip of s-SNOM also modifies the plasmon resonance energy, particularly in GNDs with small radii. To demonstrate that, we show in Fig. S3 the calculated nano-IR spectra of GND with different minimal tip-sample distance $d_0$. The radius of the GND is set to be $r = 100$ nm and $E_F$ is set to be 0.26 eV. As shown in Figure S3, the peak energy of the BM plasmon resonance decreases systematically with decreasing $d_0$. This is consistent with previous studies of graphene coupled with an adjacent metal layer (Refs. 26, 53 and 54 in the main text), where the plasmon dispersion is significantly modified. In extreme conditions, the modified graphene plasmons could have a linear dispersion, thus forming the so-called acoustic plasmons (Refs. 53 and 54 in the main text). Note that the effects of the metallic tip on the resonance energy are much smaller in larger GNDs.

### 5. Comparison of different molecular-particle shapes

In this work, we consider mainly disk-like molecular particles in our modeling for convenience. As expected, the general effects of the tip- and plasmon-enhancements of vibrational modes are qualitatively similar when using other particle shapes. As an example, we plot in Fig. S4 the calculated spectra of molecular particles with a disk-like shape versus a sphere-like shape. In both cases, the molecular particle is sitting at the center of GND with a radius of 100 nm on top of the CaF$_2$ substrate. The molecular disk has a radius of 5 nm and a thickness of 1.6 nm, and the molecular sphere has a radius of about 3.1 nm. The volume size of the two molecules is roughly the same. The doping of GND in the two cases are slightly different in order to match the resonance energy ($E_F$ is 0.26 eV and 0.23 eV for molecular disk and sphere, respectively). As shown in Fig. S4, the general spectroscopic features are similar: the vibrational mode appears as a dip at the top of the broad plasmon resonance of GND. The main difference between the two spectra lies at the intensity of both the plasmon resonance and the vibrational mode, which is mainly due to the smaller height (~50%) of the disk-shaped molecule compared to sphere-shaped molecule. A smaller molecular height indicates smaller tip-GND separation and hence stronger tip enhancement.

**Supplemental figures**

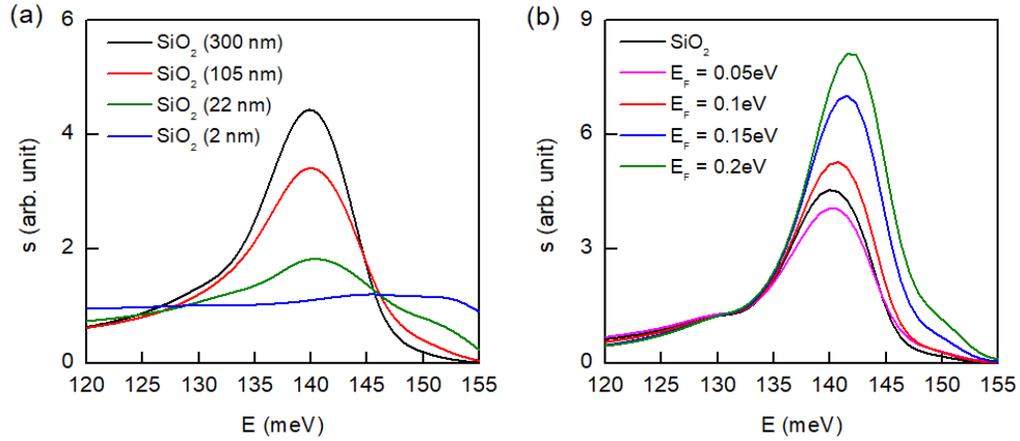

**Fig. S1.** (a) Modeled nano-IR spectra of $SiO_2$ on Si with various thickness of $SiO_2$. (b) Modeled IR amplitude spectra of large-area graphene on $SiO_2$/Si substrate versus the bare $SiO_2$/Si substrate. Thickness of $SiO_2$ here is 300 nm.

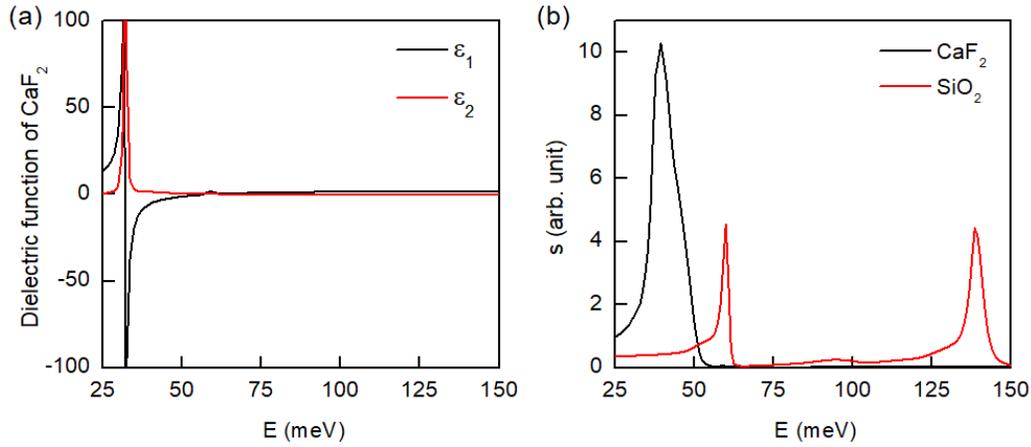

**Fig. S2.** (a) Infrared dielectric function of $CaF_2$ revealing strong phonon resonances at low energies (Ref. 47 in the main text). (b) Modeled IR amplitude spectra of $SiO_2$ and $CaF_2$ substrates normalized to that of silicon.

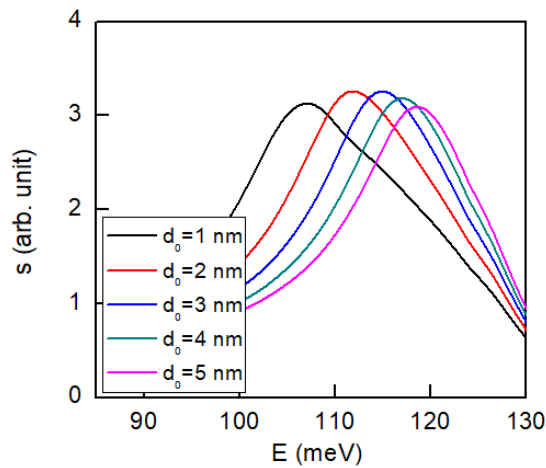

**Fig. S3.** Modeled nano-IR spectra of the BM plasmon resonance of GND with different minimal tip-sample distance ($d_0$). The spectra are normalized to that of silicon. The radius and $E_F$ of GND are set to be 100 nm and 0.26 eV, respectively.

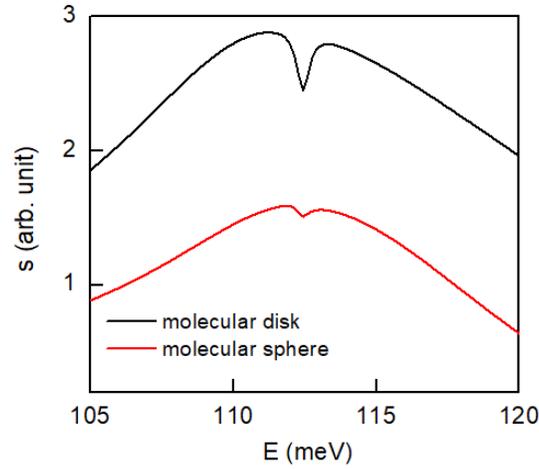

**Fig. S4.** Modeled nano-IR spectra of molecular particles with a disk shape versus a sphere shape. The spectra are normalized to that of silicon. The molecular disk has a radius of 5 nm and a thickness of 1.6 nm. The molecular sphere has a radius of about 3.1 nm. The $E_F$ of GND is 0.26 eV and 0.23 eV for the case of molecular disk and sphere, respectively.

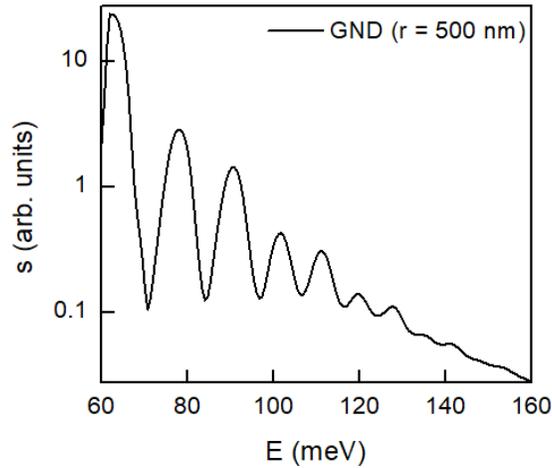

**Fig. S5.** Modeled nano-IR spectra of GNDs with a radius of 500 nm on a $CaF_2$ substrate normalized to that of silicon. The Fermi energy $E_F$ is set to be 0.2 eV. The scattering rate $\gamma$ is set to be 0.05, half of that used in the main text. With a reduced scattering rate, more resonance peaks of BM plasmons can be seen in the field of view.

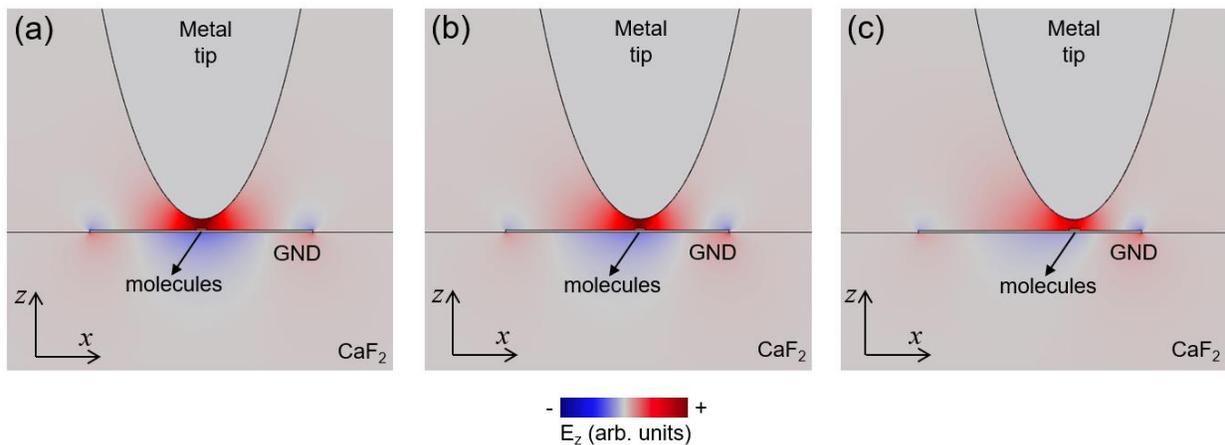

**Fig. S6.** Modeled $E_z$ field maps of the tip, molecular particle, and GND system when the tip and molecular particle are (a) at the center, (b) 20 nm away from the center, and (c) 40 nm away from the center of GND.

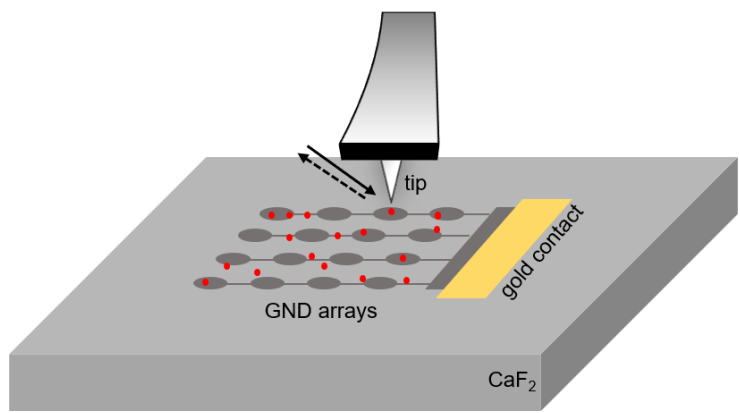

**Fig. S7.** Illustration of patterned GND arrays for the tip- and plasmon-enhanced IR spectroscopy studies of molecules (red dots) by using s-SNOM.